# Controlling Hydrogen Isotope Retention at Helium Cavities through Radiation-Induced Segregation in Fusion Steels

Lihao Shi[1] , Logan N. Clowers [1†], Qing Peng [2], Gary S. Was [1], Fei Gao[1, 3*]

*[1] Department of Nuclear Engineering and Radiological Sciences, University of Michigan, Ann Arbor, MI, 48109, USA*

*[2] School of Power and Mechanical Engineering, Wuhan University, Wuhan 430072, China*

*[3] Department of Materials Science and Engineering, University of Michigan, Ann Arbor, MI, 48109, USA*

Hydrogen isotope retention in plasma-facing and structural alloys is a central materials challenge for deuterium-tritium fusion. Motivated by ion-beam irradiation experiments and first-principles calculations, we identify an irradiation-enabled mechanism whereby solute segregation to defect sinks enhances hydrogen isotope trapping. Triple-ion irradiation of reduced activation ferritic–martensitic steel F82H reveals pronounced segregation of Cr and Ta to cavity surfaces. Density-functional theory shows that these segregants markedly increase H stability at cavities by strengthening binding energies and increasing migration barriers, leading to substantially higher attention. The effect originates from solute-tuned electronic structure: Ta promotes strong H 1s–metal d orbital hybridization, whereas Cr shifts the surface d-band toward a more favorable bonding configuration. These findings provide an atomistic link between irradiation-induced segregation and elevated-temperature hydrogen isotope retention and alloy-chemistry routes to control tritium inventory in fusion environments.

Tritium retention in structural and plasma-facing materials represents a critical challenge for the operation of future deuterium–tritium fusion reactors. Trapped tritium directly reduces fuel availability, increases operational costs, and imposes strict safety limits on the in-vessel tritium inventory. For example, ITER imposes a regulatory limit of approximately 1 kg of tritium in the vacuum vessel to mitigate radiological risks [1]. Moreover, fuel cycle modeling indicates that hydrogen isotope trapping in reactor components can significantly increase the required tritium breeding ratio (TBR) and prolong the tritium doubling time, thereby affecting the long-term sustainability of fusion power plants [2–4]. Understanding the atomic-scale mechanisms governing hydrogen isotope retention in irradiated structural materials therefore remains a central issue for fusion reactor development [5].

Hydrogen isotope retention in Fe-based structural materials is governed by a hierarchy of irradiation-induced defect sinks that differ in binding strength and thermal stability [6]. Direct cryogenic atom probe tomography (APT) has provided nanoscale evidence that hydrogen localizes at dislocations, grain boundaries, and incoherent carbide interfaces in martensitic steels [7,8], which are abundant defect sinks under irradiation conditions relevant to fusion reactors. Complementary first-principles and kinetic Monte Carlo studies further demonstrate that the atomic structure of crystalline defects governs hydrogen transport by controlling local trapping energies and diffusion pathways [9]. Complementary thermal desorption spectroscopy (TDS) measurements reveal that hydrogen trapped at dislocations is released primarily between ~100–300 °C, while trapping at grain boundaries and incoherent carbide interfaces corresponds to higher-temperature desorption peaks extending to ~300–350 °C. First-principles and thermodynamic studies further quantify the energetic hierarchy of these traps: vacancy and vacancy-cluster complexes in bcc Fe exhibit strong hydrogen binding energies of ~0.5–0.6 eV [10,11], while hydrogen binding at Fe–He interfaces ranges from ~0.59–0.92 eV depending on surface orientation and local helium density, although the binding strength generally decreases with increasing interfacial helium concentration [12].

More recently, Clowers et al. [13,14] employed controlled triple-ion beam irradiations to decouple the respective roles of displacement damage, helium, and hydrogen under fusion-relevant conditions. Across multiple Fe-based alloys irradiated between 400 and 600 °C, co-injection of hydrogen with helium consistently increased both cavity density and cavity size compared with He-only irradiation. Electron energy-loss spectroscopy (EELS) further revealed a distinct core–shell structure at 450 °C in Fe–Cr–W alloys, where helium preferentially occupies the cavity interior while hydrogen segregates near the cavity surface. Complementary molecular dynamics (MD) simulations suggested that hydrogen can be transiently trapped at bubble interfaces, but becomes thermally unstable at elevated temperatures, with de-trapping occurring above ~127 °C [15,16]. The calculated de-trapping enthalpy is ~0.703 eV [17], in good agreement with thermal desorption spectroscopy (TDS) measurements in He-implanted Fe that reported hydrogen/deuterium trapping energies of 0.75 eV, corresponding to de-trapping temperatures near 227–327 °C [18]. These observations imply that hydrogen retention near helium cavities is strongly temperature dependent and cannot yet be fully explained by current atomistic models.

---

†Current affiliation: Materials Science and Technology Division, Oak Ridge National Laboratory, Oak Ridge, Tennessee 37831, USA.
*Contact author: gaofeium@umich.edu

However, most atomistic studies have focused on pure Fe, whereas fusion structural materials such as RAFM steels contain multiple alloying elements, and their interactions with irradiation-induced defect sinks, including voids and helium cavities, remain largely unexplored. Solutes can modify hydrogen adsorption and defect energetics through electronic and chemical effects [19–21], and chromium in particular may preferentially segregate to helium bubble interfaces under compressive environments [22]. These coupled interactions between helium, radiation-induced segregation, and alloy chemistry may therefore critically influence hydrogen isotope retention in fusion environments, yet they remain poorly understood.

To address these puzzles, we combine experiments and computer simulations to study how composition and chemical inhomogeneity affect hydrogen interactions with cavities. Experimental characterization was performed on triple-ion–irradiated F82H-IEA alloy using scanning transmission electron microscopy combined with energy dispersive X-ray spectroscopy (STEM-EDS). Details of the sample preparation and irradiation procedures have been reported previously [13,14,23] and are summarized here briefly; the detailed alloy compositions are listed in Table S1. The irradiated materials were produced using the triple-beam capabilities at the Michigan Ion Beam Laboratory (MIBL), where Fe, H, and He ions were simultaneously implanted to reproduce irradiation conditions relevant to fusion environments. Specifically, 5 MeV $Fe^{2+}$, 390 keV $H^{+}$, and 2.85 MeV $He^{2+}$ ion beams were used to generate displacement damage together with concurrent hydrogen and helium implantation. Irradiations were conducted at 500 °C to doses of 50 and 150 dpa with implantation rates of approximately 40 and 10 appm $dpa^{-1}$ for H and He, respectively.

Representative EDS maps of cavities formed after irradiation to 50 and 150 dpa are shown in Fig. 1 and Fig. S1, respectively. Each set of images reveals clear segregation of chromium and tantalum at cavity surfaces throughout the depth-dependent damage region, while tungsten does not exhibit pronounced segregation but remains broadly distributed around the cavity region. In the same samples, hydrogen enrichment at He cavity surfaces was also experimentally observed and reported in our previous study [23], where a pronounced H-rich shell was found to surround the cavity interface. Together, these observations demonstrate that helium cavities in irradiated RAFM steels form chemically complex interfaces enriched with both alloying elements and hydrogen, rather than idealized pure Fe surfaces.

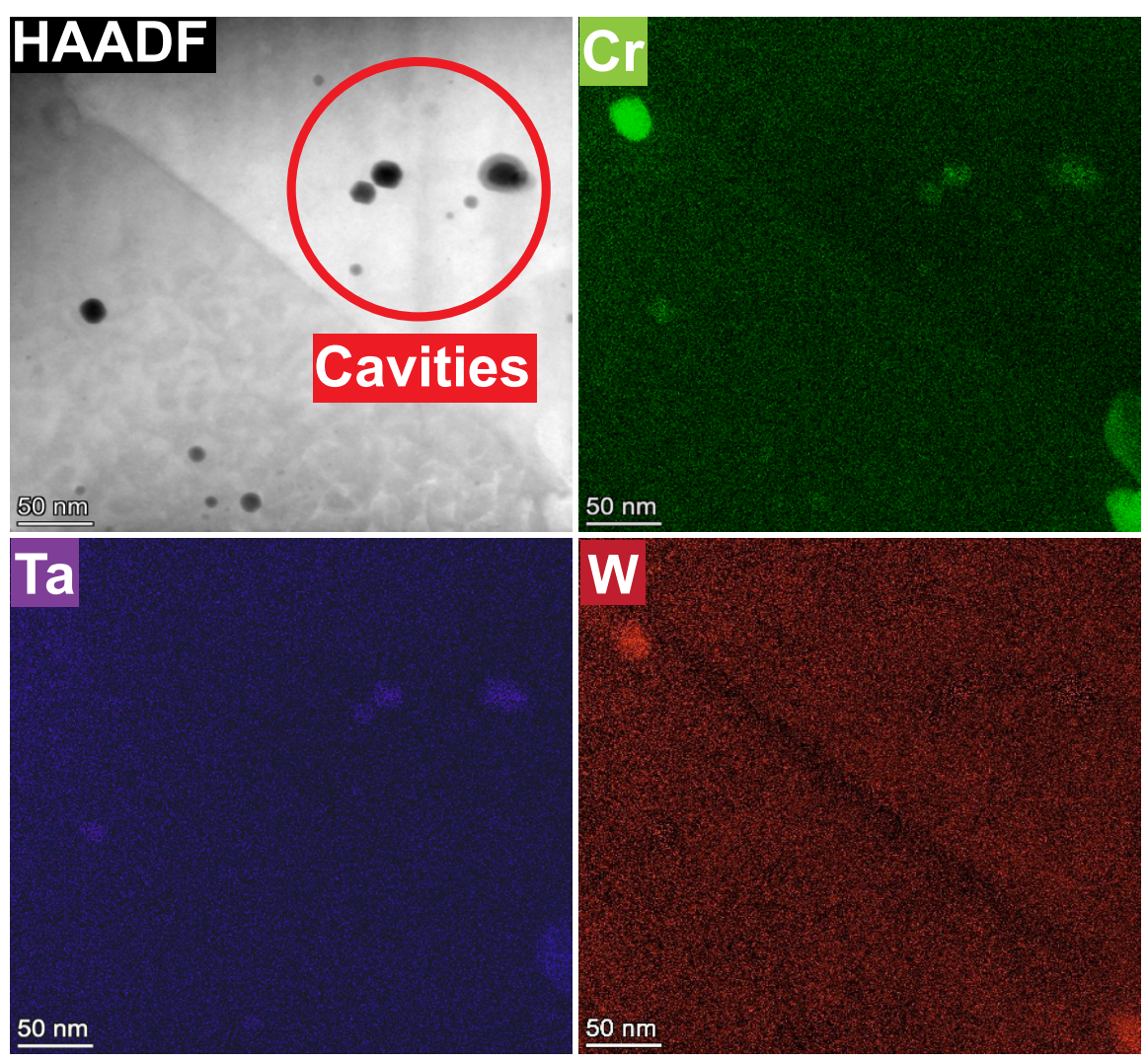

FIG. 1. EDS maps (in wt%) of chromium, tantalum taken from F82H-IEA triple ion irradiated at 500°C to 50dpa , each at 40/10 appm/dpa of H/He.

Guided by these observations, we investigated H stability at Cr/Ta/W-enriched interfaces by determining electronic structures and chemical inhomogeneity, and by calculating H binding energy at both void and He cavity surfaces. By examining the effects of helium density and solute coverage on hydrogen stability and migration barriers, this study establishes an atomistic framework linking solute segregation with hydrogen retention at helium cavities. The H binding energy ( $E_b$ ) is calculated as the difference between the H solution energy at a tetrahedral interstitial site in bulk Fe and the H adsorption energy at the cavity surface. The H adsoption energy was calculated as $E_{ads}^{H} = E_{slab}^{He+H} - E_{slab}^{He} - \frac{1}{2}E_{H_2}$ . All first-principles calculations were performed using the Vienna ab initio Simulation Package (VASP) [24]. Because experiments observed that helium bubble facets predominantly align with {100} planes in bcc Fe [25] and theoretical studies showed strong hydrogen adsorption on Fe (110) surfaces [26–28], two representative surface orientations, (100) and (110), were considered, as shown in Figs. 2(a) and (b). Helium cavities were modeled by introducing ordered helium clusters above Fe surfaces, with densities corresponding to He-to-vacancy (He/V) ratios of 0.5 and 1, reflecting the liquid-like, density-governed nature of helium assemblies [22,29]. Substitutional alloying elements (Cr, Ta and W) were introduced into the top Fe layer with coverages ranging from 0.1. 0.5 and 1 ML and hydrogen was initially placed at several candidate adsorption sites as shown in Figs. 2(c) and (d).

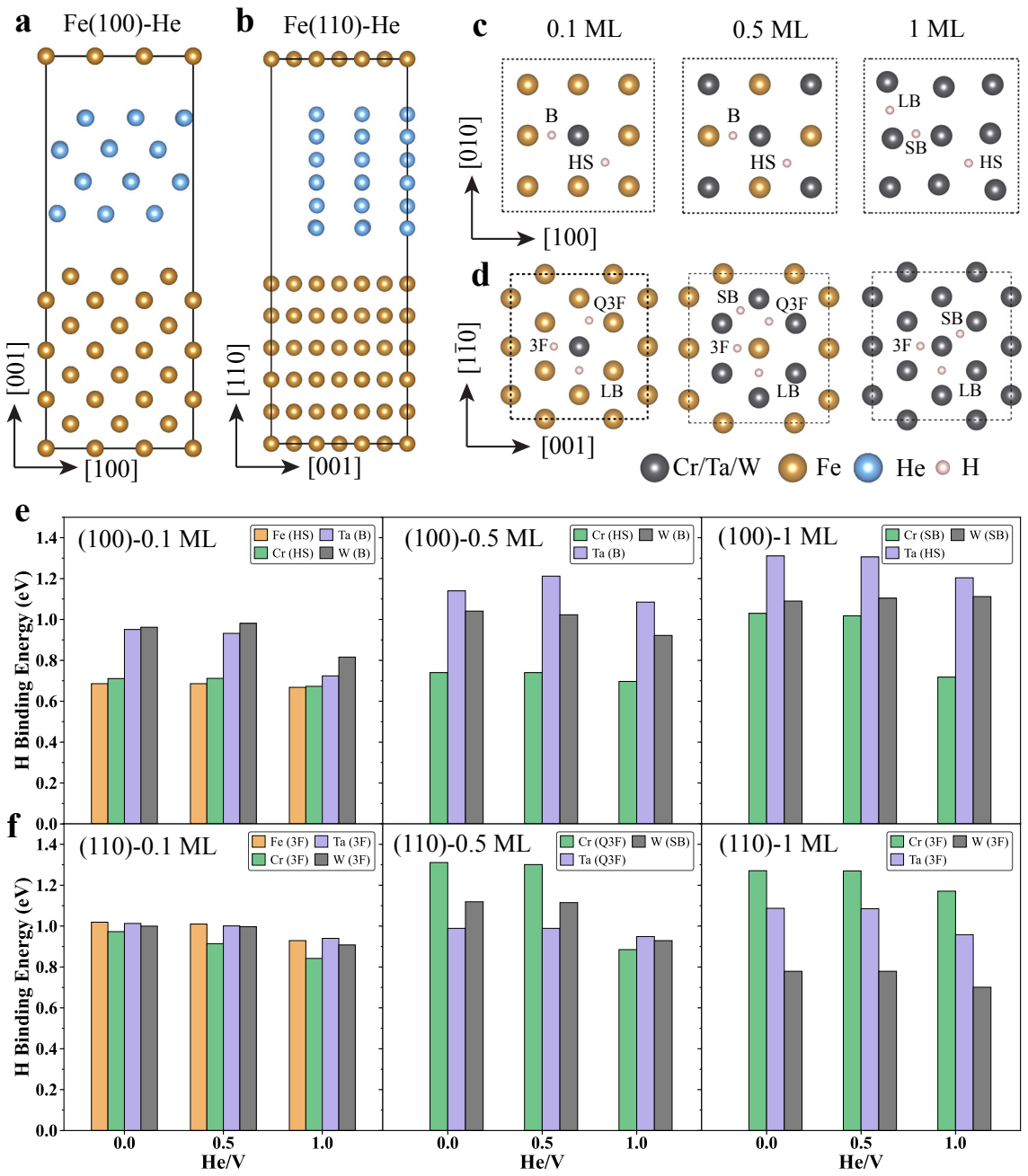

FIG. 2. DFT models and H binding energies at solute-enriched Fe–He cavity interfaces. (a, b) Fe(100) and Fe(110) surfaces containing ordered He clusters. (c,d) H adsorption sites considered on the (100) and (110) surfaces, respectively. (e, f) H binding energies at the most stable adsorption sites as a function of He/V ratio and Cr, Ta, and W coverage.

The H binding energies at the most stable adsorption sites are summarized in Figs. 2(e, f). On both Fe(100) and Fe(110) surfaces, segregation of Cr, Ta, and W generally enhances H binding relative to pure Fe. The magnitude of this enhancement depends strongly on surface orientation and solute coverage. On Fe(100), Ta produces the largest increase in H binding energy, whereas on Fe(110), Cr becomes increasingly favorable at higher coverages. In contrast, helium has a weaker effect. A moderate helium density (He/V = 0.5) results in little change compared with the corresponding void surface (He/V = 0), while a higher helium density (He/V = 1) consistently reduces H binding strength. A similar helium-density dependence is observed on Fe(110), although the coverage dependence becomes strongly element-specific. Tungsten exhibits maximum H trapping at intermediate coverages, whereas full monolayer coverage reduces the H binding energy to values comparable to or lower than those of pure Fe. In addition to enhancing thermodynamic stability, solute segregation also increases the kinetic stability of hydrogen at cavity interfaces. CI-NEB calculations (Fig. S2) show that alloying elements generally increase the migration barrier for H escape from the interface, consistent with the corresponding increase in H binding energy. On the Fe(100) surface, Ta produces the largest barrier enhancement, whereas Cr has a stronger effect on Fe(110). Increasing helium density tends to reduce the migration barrier, particularly for pure Fe and Cr-decorated surfaces, reflecting the destabilizing effect of over-pressurized helium on interfacial hydrogen. In contrast, the migration barrier on Ta-decorated surfaces is comparatively insensitive to helium accumulation, indicating that Ta remains effective in stabilizing hydrogen even under high-He conditions. Together, these results demonstrate that hydrogen retention at cavity interfaces is strongly influenced by both local composition and helium density, motivating a detailed examination of the underlying electronic mechanisms governing H stability.

The origin of solute-modified hydrogen stability at cavity surfaces is summarized in Fig. 3. The first contribution arises from the local chemical identity of the alloying elements, which directly modifies metal–H bonding. As shown in Fig. 3a, for the single-solute cases with H adsorbed at the bridge (B) site on Fe(100), Ta and W atoms enhance hydrogen stability through strong hybridization between the solute d orbitals and the H 1s orbital. This effect is evident from the projected crystal orbital Hamilton population (COHP) analysis [30], where Ta–H and W–H bonds exhibit substantially larger bonding contributions than Fe–H. Consistently, the integrated crystal orbital Hamilton population (-ICOHP), which quantifies the overall bond strength, correlates positively with the H binding energy. These results indicate that stronger local metal–H bonding promotes enhanced hydrogen stability at the cavity interface.

Increasing solute coverage further modifies H stability by altering both the local bonding geometry and the electronic environment. On the pure Fe(100) surface, H occupies a symmetric hollow site (HS) coordinated by neighboring Fe atoms. In contrast, when a Ta monolayer is present at the interface, H becomes stabilized near the surface Ta atoms, as illustrated by the atomic configurations in Fig. 3b. The corresponding charge-density-difference maps reveal substantial electron accumulation around the adsorbed

H, together with charge depletion around neighboring metal atoms, indicating enhanced charge transfer to H. Quantitative Bader analysis further confirms this trend [31], showing a positive correlation between the charge transferred to H and the H binding energy for (100) interfaces (Fig. 3c). These results demonstrate that charge transfer serves as an effective descriptor of solute-enhanced hydrogen stabilization on the (100) surface.

hollow sites (HS) for (100) interfaces and threefold sites (3F) for (110) interfaces. Here, $E_{ads}^{chem}$ denotes the chemical component of the H adsorption energy ($E_{ads}^{H}$) used to determine the H binding energies in Fig. 2, with the structural relaxation contribution removed through the frozen-surface approach. The resulting chemical adsorption energies show a clear correlation with the surface d-band center, as shown in Fig. 3e. The corresponding H binding energies are also plotted on the right axis for comparison.

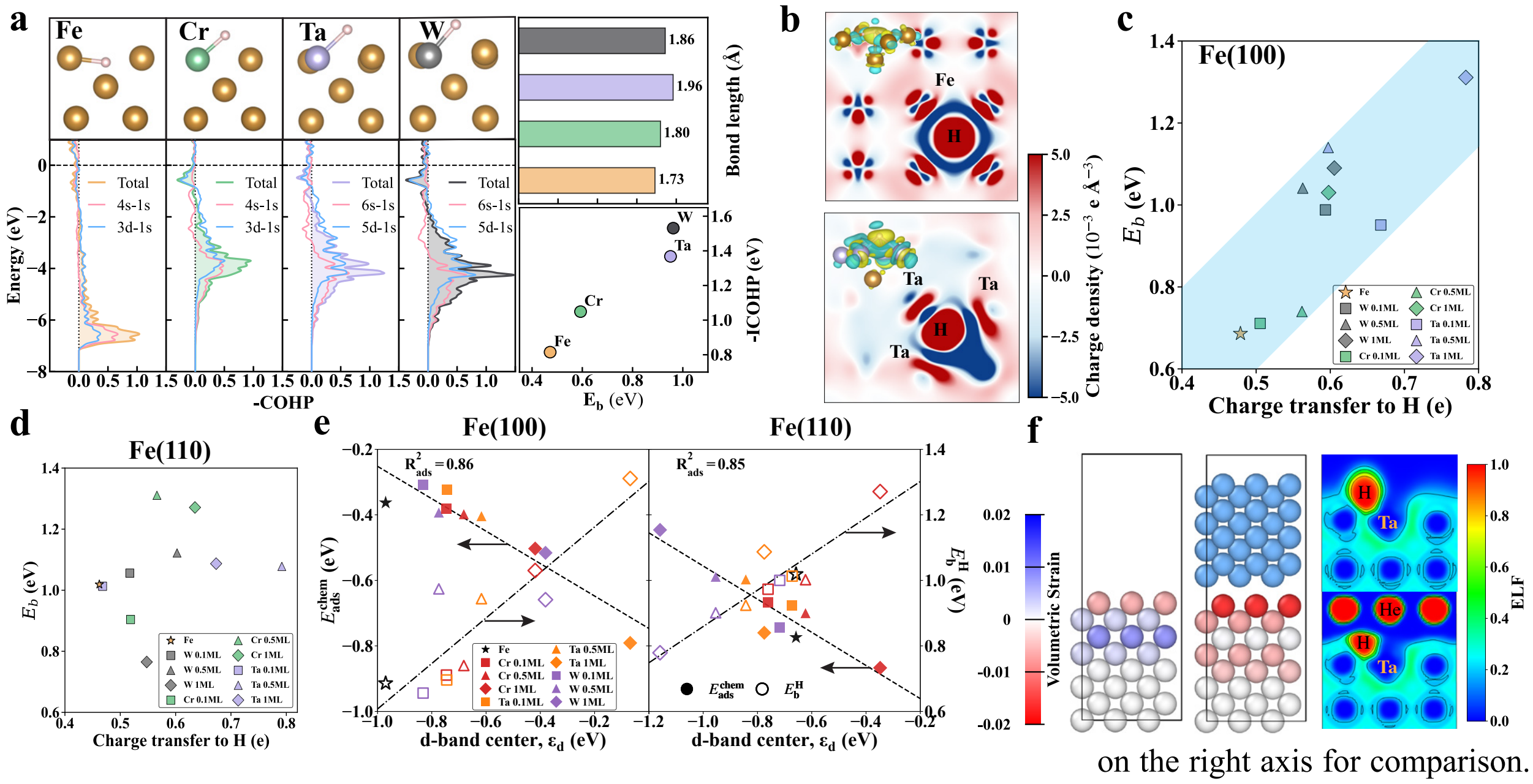

FIG. 3. (a) COHP, bond lengths, and -ICOHP for H bonded to Fe, Cr, Ta, and W on Fe(100). (b) Charge density difference maps for H adsorption on pure and Ta-decorated Fe(100) surfaces. Yellow and cyan isosurfaces indicate electron accumulation and depletion, respectively. (c,d) H binding energy versus charge transferred to H for Fe(100) and Fe(110), respectively. (e) Chemical H adsorption energy (left axis) and the corresponding H binding energy (right axis) as a function of the surface d-band center. (f) Volumetric strain and electron localization function (ELF) around H at He/V = 0 and 1 cavity interfaces.

While charge transfer provides an effective descriptor of H stabilization on Fe(100) interfaces, the electronic factors governing H adsorption extend beyond local charge redistribution alone. As shown in Fig. 3d, the dependence of H binding energy on charge transfer becomes weaker for Fe(110) interfaces, particularly for Ta and W-decorated surfaces. Although the supplementary COHP analysis (Fig. S3) indicates that W and Ta continue to form strong local metal–H bonds on the Fe(110) surface, increasing W coverage to 1 ML reduces the overall H binding energy to values below those of the pure Fe surface. These observations indicate that, in addition to local bonding and charge transfer, H stability is also influenced by the collective electronic structure of the surface.

To capture this effect, we evaluated the chemical component of H adsorption using frozen-surface calculations [32]: $E_{ads}^{chem} = E_{slab}^{H} - E_{slab}^{frozen} - \frac{1}{2}E_{H_2}$, This correlation is strong for both Fe(100) and (110) surfaces, demonstrating that solute-induced redistribution of surface d states controls the intrinsic chemical driving force for H adsorption. Within the d-band framework, an upward shift of the surface d-band center favors H adsorption by enhancing bonding-state occupation while keeping antibonding states less occupied. In contrast, a downward shift increases antibonding-state occupation and weakens the chemical contribution to H adsorption. Consistent with this picture, W significantly lowers the d-band center on Fe(110) surface, reducing the chemical contribution to H binding despite the presence of strong local W–H interactions. By contrast, a Cr monolayer shifts the surface d-band center toward the Fermi level, thereby promoting stronger H adsorption. Therefore, hydrogen stability at cavity interfaces is governed by the synergistic effects of local solute–H hybridization, charge transfer, helium-induced strain (see below), and the global surface electronic structure. Together, these factors determine both the local bonding environment and the overall chemical driving force for H adsorption.

Helium accumulation further modifies hydrogen stability through coupled mechanical and electronic perturbations at the cavity interface. As shown in Fig. 3f for Fe(100) and Fig. S4 for Fe(110), increasing helium density generates a strong compressive strain field near the cavity surface, consistent with previous studies of over-pressurized helium bubbles [26]. A similar response is observed for both surface orientations, indicating that the effect originates primarily from the local helium environment rather than the specific surface geometry. The corresponding electron localization function (ELF) maps reveal a pronounced compression of the H localization basin in the presence of helium, reflecting a reduced ability of neighboring metal atoms to transfer charge to H. Quantitative Bader analysis further supports this interpretation, showing a reduced H charge gain at He/V = 1 compared with the helium-free interfaces (Fig. S5). Consequently, the reduced charge transfer provides an electronic origin for the lower H binding energies observed at over-pressurized helium cavities. Together, these results demonstrate that helium weakens hydrogen trapping through coupled mechanical and electronic effects associated with helium-induced compression.

This study uncovers atomistic mechanisms for enhanced hydrogen retention at helium-cavity interfaces in candidate fusion blanket materials. Guided by experimental observations of Cr, Ta and W segregation at irradiation-induced cavities, first-principles calculations show that solute-enriched interfaces exhibit stronger hydrogen trapping and higher migration barriers than pure Fe surfaces, thereby enhancing hydrogen retention both thermodynamically and kinetically. Electronic-structure analysis reveals that the enhanced stability originates from the combined effects of strengthened local metal–H hybridization, increased charge transfer to hydrogen, and solute-induced modification of the surface d-band structure, which collectively increase the intrinsic chemical driving force for H adsorption. In contrast, helium accumulation introduces a destabilizing effect through helium-induced compression, which reduces charge transfer and weakens hydrogen stabilization at over-pressurized cavity interfaces. Nevertheless, Cr- and Ta-segregated interfaces remain effective hydrogen traps over a wide range of helium densities, explaining the experimentally observation of high-temperature hydrogen retention at both void and helium-cavity surfaces.

***Acknowledgments***— This work was supported by the U.S. Department of Energy, Office of Science, Office of Fusion Energy Sciences under Award Number DE-SC0020226. This research used resources of the Oak Ridge Leadership Computing Facility at the Oak Ridge National Laboratory, which is supported by the Office of Science of the U.S. Department of Energy under Contract No. DE-AC05-00OR22725. The authors also gratefully acknowledge Zhijie Jiao, Ovidiu Toader, Fabian Naab, Thomas Kubley, Robert Hensley, Catherine Nicoloff, Alexander Flick and Prashanta Niraula at the Michigan Ion Beam Laboratory for their assistance with the ion irradiations, the Michigan Center for Materials Characterization for use of the instruments used in this study, and Hiroyasu Tanigawa at the National Institutes for Quantum and Radiological Science and Technology (QST) in Aomori, Japan for providing the F82H-IEA alloy.

***Data availability***—The data supporting the findings of this article are available upon reasonable request.